\documentclass[twocolumn,showpacs,preprintnumbers,amsmath,amssymb]{revtex4}
\usepackage{graphics}
\usepackage{graphicx}

\begin{document}

\title{Fidelity, dynamic structure factor, and susceptibility in critical phenomena}
\author{Wen-Long You$^{1,2}$} \author{Ying-Wai Li$^{1}$} \author{Shi-Jian Gu$^1$}
\altaffiliation{Email: sjgu@phy.cuhk.edu.hk} \affiliation{$^1$Department of
Physics and Institute of Theoretical Physics, The Chinese University of Hong
Kong, Hong Kong, China\\$^2$School of Physics, Peking University, Beijing,
100871, China}
\date{\today}

\begin{abstract}
Motivated by the growing importance of fidelity in quantum critical phenomena,
we establish a general relation between fidelity and structure factor of the
driving term in a Hamiltonian through a newly introduced concept:
fidelity susceptibility. Our discovery, as shown by some examples, facilitates the
evaluation of fidelity in terms of susceptibility using well developed techniques
such as density matrix renormalization group for the ground state, or Monte Carlo
simulations for the states in thermal equilibrium.
\end{abstract}




\maketitle

\section{Introduction}

Recently, much attention
\cite{HTQuan2006,Pzanardi2006,Pzanardi0606130,Pzanardi0612006,PBuonsante0612590,HQZhou07}
has been drawn to the role of fidelity, a concept emerged from quantum
information theory \cite{Nielsen1}, in quantum critical phenomena
\cite{Sachdev}. Since fidelity is a measure of similarity between states, a
dramatic change in the structure of the ground state around quantum critical
point should result in a great difference between the two ground states on the
both sides of critical point. For example, in one-dimensional XY model, fidelity
shows a narrow trough at phase transition point \cite{Pzanardi2006}. Similar
properties were also found in fermionic \cite{Pzanardi0606130} and bosonic
systems \cite{PBuonsante0612590}. As fidelity is purely a quantum information
concept, these works actually established a connection between quantum
information theory and condensed matter physics.

However, except a few specific models such as the one-dimensional XY model and
Dicke model\cite{HTQuan2006,Pzanardi2006}, it is tedious to evaluate fidelity
from the ground state wavefunctions. Therefore, a neater and simpler formalism
is of great importance for the extensive application of fidelity to critical
phenomena. For this purpose, we introduce the concept of fidelity susceptibility,
which defines the response of fidelity to the driving parameter of the Hamiltonian.
At zero temperature, we show that fidelity susceptibility is intrinsically related
to the dynamic structure factor of the driving Hamiltonian, namely $H_I$, that
causes quantum phase transition. Based on some well-developed numerical techniques
for the ground state properties such as exact diagonalization (ED) \cite{ED}
and density matrix renormalization group (DMRG) \cite{DMRG}, an applicable scheme
is proposed to evaluate the dynamic structure factor of $H_I$. On the other hand,
starting from the definition of fidelity of a thermal state, we show that fidelity
susceptibility is simply the thermal fluctuation term, such as specific heat $C_v$
for the internal energy and magnetic susceptibility $\chi$ for magnetization.
These can easily be obtained from Monte-Carlo simulations \cite{DPLandaub}.

\section{Ground-state fidelity and dynamic structure factor}

A general Hamiltonian of quantum many-body systems reads

\begin{eqnarray}
H(\lambda)=H_0 +\lambda H_I, \label{eq:Hamitonian}
\end{eqnarray}
where $H_I$ is the driving Hamiltonian and $\lambda$ denotes its strength. The eigenstates
$|\Psi_n(\lambda)\rangle$ which satisfy $H(\lambda)|\Psi_n(\lambda)\rangle =E_n |\Psi_n(\lambda)\rangle$
define a set of orthogonal complete basis in the Hilbert space. Here we restrict ourselves
to the phase transition which is not induced by the ground state level-crossing. That means the
ground state of the Hamiltonian is non-degenerate for a finite system. Next we change
$\lambda\rightarrow\lambda+\delta\lambda$ where $\delta\lambda$ is so small that perturbation
is applicable. To the first order, the ground state becomes

\begin{eqnarray}
|\Psi_0(\lambda+\delta\lambda)\rangle=|\Psi_0(\lambda)\rangle +\delta\lambda
\sum_{n\neq 0} \frac{H_{n0}(\lambda)
|\Psi_n(\lambda)\rangle}{E_0(\lambda)-E_n(\lambda)} \label{eq:perturbatios}
\end{eqnarray}
and
\begin{eqnarray}
H_{n0}=\langle\Psi_n(\lambda)|H_I|\Psi_0(\lambda) \rangle.
\end{eqnarray}
Following Ref. \cite{Pzanardi2006}, fidelity is defined as the overlap between
$|\Psi_0(\lambda)\rangle$ and $|\Psi_0(\lambda+\delta\lambda)\rangle$, that is

\begin{eqnarray} F_i (\lambda,
\delta)=|\langle\Psi_0(\lambda)|\Psi_0(\lambda+\delta)\rangle|.
\label{eq:fidedifintion}
\end{eqnarray}
Therefore, to the lowest order, we have

\begin{eqnarray}
{F_i^2}=1 - \delta\lambda^2 \sum_{n\neq
0}\frac{|\langle\Psi_n(\lambda)|H_I|\Psi_0(\lambda)
\rangle|^2}{[E_n(\lambda)-E_0(\lambda)]^2 } +\cdots. \label{eq:fidelityexp}
\end{eqnarray}
Clearly, fidelity is $\delta\lambda$-dependent and so it is an artificial quantity.
Despite of this, we can still see from Eq.(\ref{eq:fidelityexp}) that the most
relevant term in determining fidelity is its second derivative. Compared with linear
response theory, the coefficient term before $\delta\lambda^2$ actually defines
the response of fidelity to a small change in $\lambda$. From this point of view,
we introduce a new concept \emph{fidelity susceptibility} as

\begin{eqnarray}
\chi_F\equiv \lim_{\delta\lambda\rightarrow 0}\frac{-2\ln
F_i}{\delta\lambda^2}.
\end{eqnarray}
With Eq. (\ref{eq:fidelityexp}), it can be rewritten as

\begin{eqnarray}
\chi_F(\lambda)=\sum_{n\neq 0}\frac{|\langle\Psi_n(\lambda)|H_I|\Psi_0(\lambda)
\rangle|^2}{[E_n(\lambda)-E_0(\lambda)]^2} \label{eq:Fidelityexp2}
\end{eqnarray}
in the ground state. We would like to point out that although the aforementioned
procedure is based on perturbation theory, fidelity susceptibility
(\ref{eq:Fidelityexp2}) only depends on the spectra of the Hamiltonian
$H(\lambda)$ and the hopping matrix $H_{n0}$. Unfortunately, except for some
very small systems which are usually far away from the scaling region,
Eq.(\ref{eq:Fidelityexp2}) is almost not computable due to the lack of knowledge
in the set of eigenstates. In order to overcome the difficulty, it is necessary
to consider the time evolution of the system. For simplicity, we omit the
parameter $\lambda$ in the following expressions. Define \emph{dynamic fidelity
susceptibility} as

\begin{eqnarray}
\chi_F(\omega)=\sum_{n\neq 0}\frac{|\langle\Psi_n|H_I|\Psi_0
\rangle|^2}{[E_n-E_0]^2 + \omega^2}
\end{eqnarray}

Perform a Fourier transformation and take derivative, we then obtain

\begin{widetext}
\begin{eqnarray}
\frac{\partial\chi_F(\tau)}{\partial \tau} = -\pi\left[\langle \Psi_0 |H_I(\tau)
H_I(0)|\Psi_0\rangle -\langle\Psi_0|H_I|\Psi_0\rangle^2 \right]\theta(\tau)+
\pi\left[\langle \Psi_0 |H_I(0) H_I(\tau)|\Psi_0\rangle
-\langle\Psi_0|H_I|\Psi_0\rangle^2 \right]\theta(-\tau),
\label{eq:fed_fluctuation}
\end{eqnarray}
\end{widetext}
with $\tau$ being the imaginary time and
\begin{eqnarray}
H_I(\tau)=e^{H(\lambda)\tau} H_I e^{-H(\lambda)\tau}.
\end{eqnarray}
The two equations mentioned above are impressive as they reveal the
mystery of fidelity in understanding quantum critical phenomena. The
terms in the brackets in Eq. (\ref{eq:fed_fluctuation}) are nothing
but the dynamic structure factor of $H_I$. Therefore in the original
definition of fidelity, we subconsciously choose the driving term
$H_I$ as a candidate of the order parameter, though we might not
think so at that time.

In order to arrive at a more computable formula, we carry out an inverse Fourier
transformation and obtain

\begin{eqnarray}
\chi_F=\int\tau\left[ \langle \Psi_0 |H_I(\tau) H_I(0)|\Psi_0\rangle
-\langle\Psi_0|H_I|\Psi_0\rangle^2\right] d\tau \nonumber \\
\label{eq:fidelityfnal}
\end{eqnarray}
where the first term in the bracket can be calculated by

\begin{eqnarray}
&&\langle \Psi_0 |H_I(\tau) H_I(0)|\Psi_0\rangle \nonumber \\ && = \sum_n
\frac{\tau^n (-1)^n}{n!}e^{\tau E_0}\langle \Psi_0 |H_I H^n H_I|\Psi_0\rangle.
\label{eq:dyfactor}
\end{eqnarray}
Although fidelity is difficult to calculate from the ground state wavefunctions,
Eq. (\ref{eq:fidelityfnal}) and (\ref{eq:dyfactor}) provide us with another
practical way. Especially, Eq. (\ref{eq:dyfactor}) can be easily evaluated via
the prevailing numerical techniques, say, ED and DMRG. For the ED, once the
ground state is obtained, the map from one state to another new state is just a
standard Lanczos step. While for the DMRG, the standard algorithm involves a
transformation of the Hamiltonian of a system and its environment, from a set of
old basis to a set of new basis which is constructed by the $m$ largest weighted
eigenstates of the reduced density matrix. Precisely, for the system block
$\overline{H}_L=O_L^\dagger H_L O_L$, and environment block
$\overline{H}_R=O_R^\dagger H_R O_R$, where $O_{L(R)}$ are constructed from $m$
largest weighted eigenstates of the corresponding reduced density matrix. The
only modification is that, in addition to $H(\lambda)$, $H_I$ should be
independently transformed in the DMRG procedure, i.e.
$\overline{H}_{I,L}=O_L^\dagger {H}_{I,L} O_L$ and
$\overline{H}_{I,R}=O_R^\dagger {H}_{I,R} O_R$. Once the final ground state is
obtained, the mapping $|\Psi'\rangle=H_I|\Psi\rangle$ and
$|\Psi'\rangle=H|\Psi\rangle$ is simply a standard step.

\begin{figure}
\includegraphics[width=8cm]{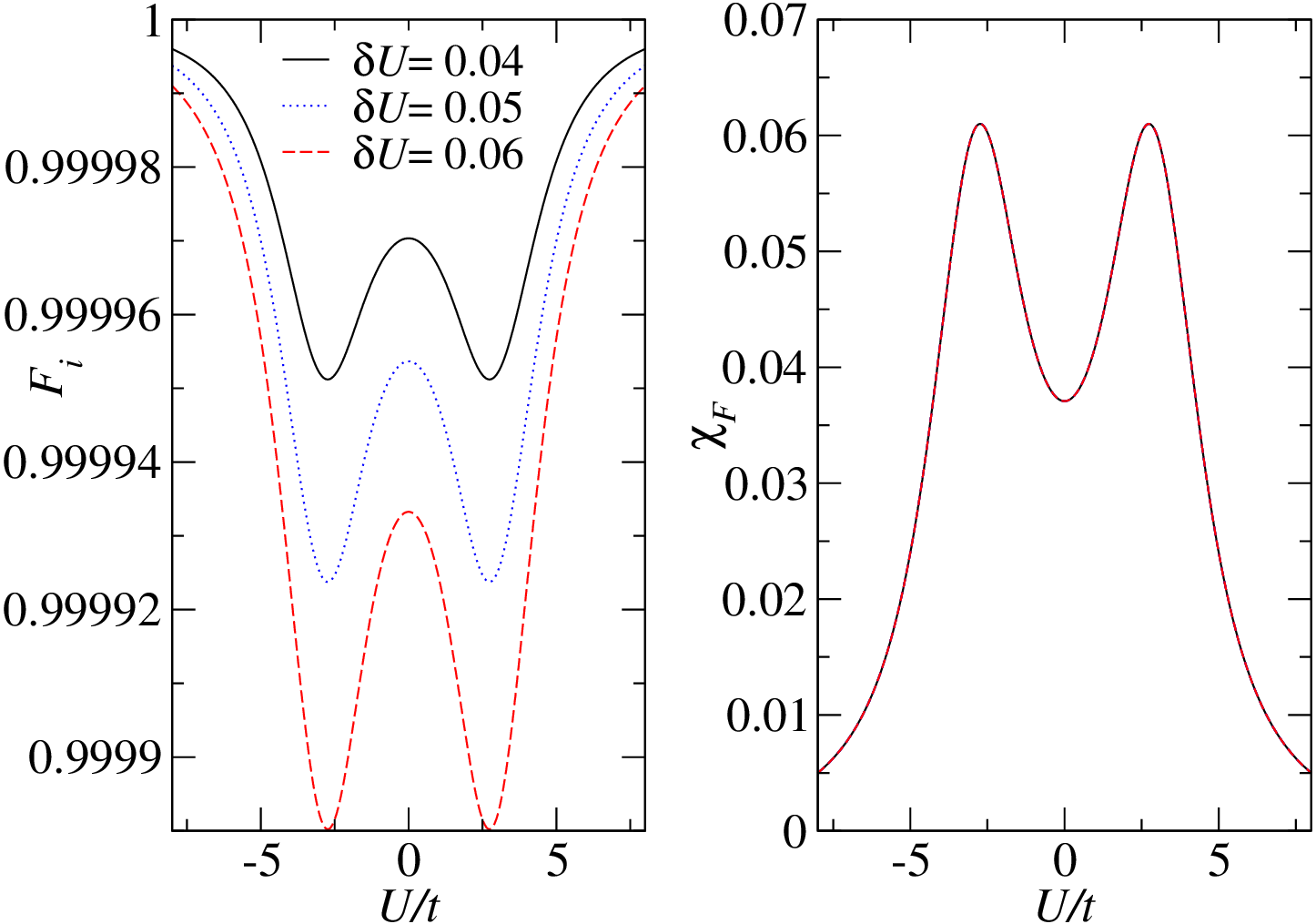}
\caption{(Color online) LEFT: the fidelity between two states separated by
different $\delta U=0.04, 0.05, 0.06$, versus $U/t$ of the half-filled Hubbard
model with $N=L=10$ and periodic boundary conditions. RIGHT: the fidelity
susceptibility $\chi_F$ as a function of $U/t$, obtained from the data of the
left picture. All lines in the left picture collapse to a single line.
\label{fig:fid_hub}}
\end{figure}

\begin{figure}
\includegraphics[width=8cm]{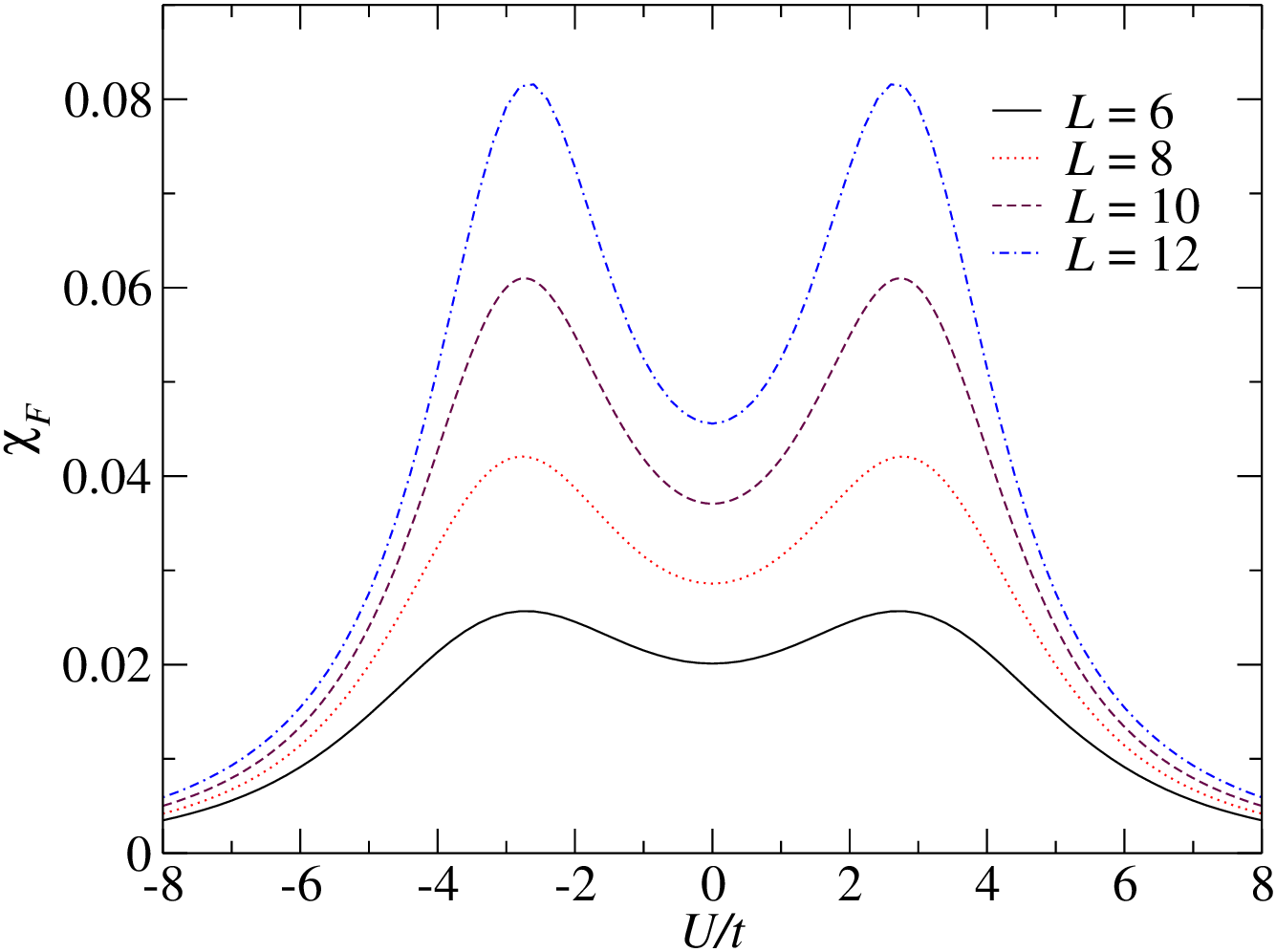}
\caption{(Color online) Fidelity susceptibility as a function of $U/t$ for
various system size. In order to avoid degeneracy in the ground state, the
systems of $L=6$ and $L=10$ are diagonalized with periodic boundary conditions,
while the systems of $L=8$ and $L=12$ with antiperiodic boundary conditions.
\label{fig:fid_hubscale}}
\end{figure}

\begin{figure}
\includegraphics[width=7cm]{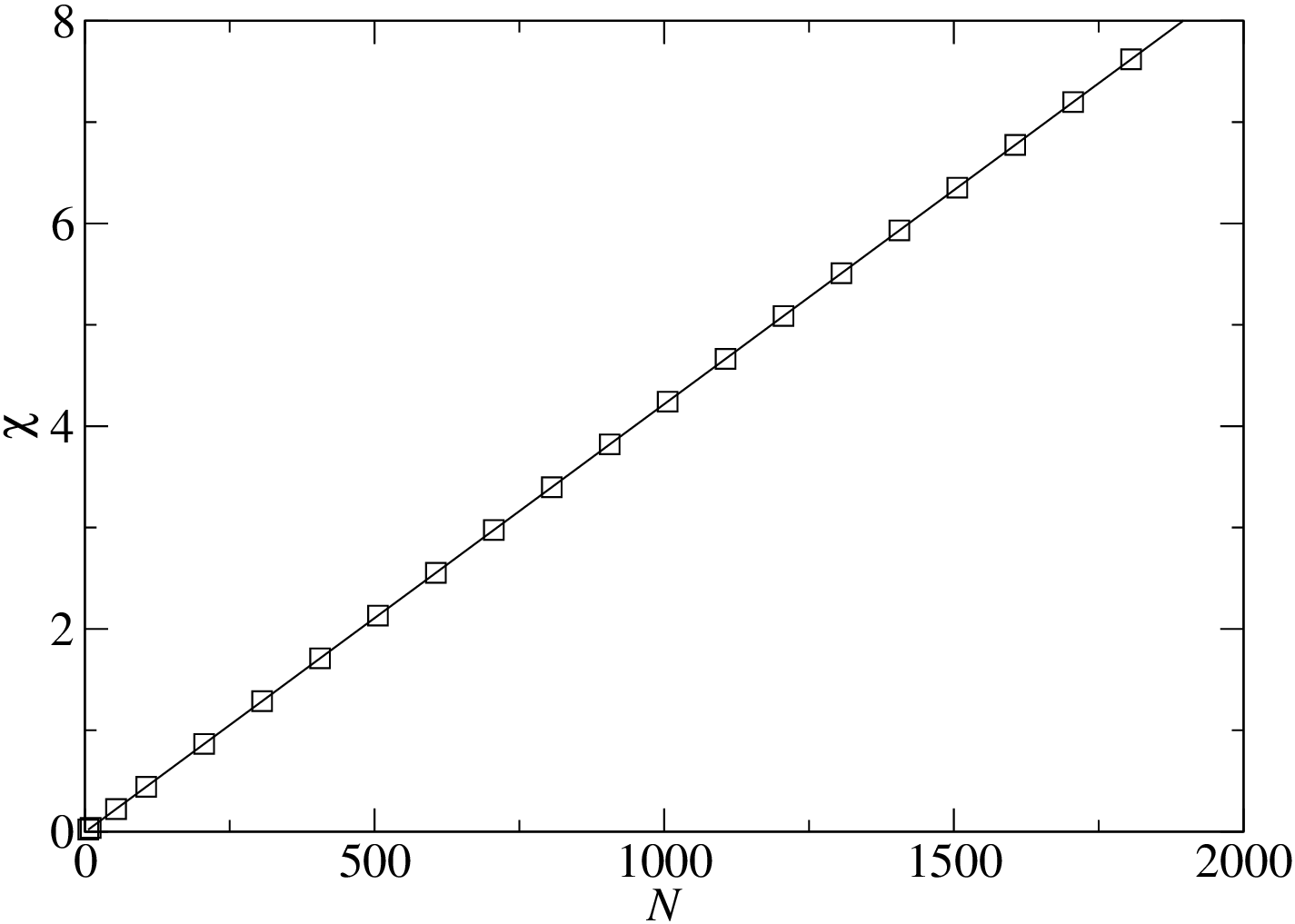}
\caption{The scaling behavior of the fidelity susceptibility up to
1906 sites for the half-filled Hubbard at $U=0$. Clearly, $\chi/N$
is a constant if $N$ is very large. \label{fig:chi_hub}}
\end{figure}

\section{Fidelity and fidelity susceptibility in some models}

In previous works
\cite{HTQuan2006,Pzanardi2006,Pzanardi0606130,Pzanardi0612006,PBuonsante0612590,HQZhou07},
the fidelity gives us a deep impression that it can be used to
signal any phase transition. For example, in the Ising
model,\cite{HTQuan2006,Pzanardi2006} the fidelity collapses to zero
at the critical point. From Eq. (\ref{eq:fidelityexp}), such a
singular behavior can be interpreted as the divergence of the
fidelity susceptibility. This interpretation is consistent with that
the fidelity is intrinsically based on the Landau's
symmetry-breaking theory. The order parameter is just the driving
term in the Hamiltonian, then the divergence of the fidelity
susceptibility becomes a crucial condition to signal a phase
transition of Landau's type. However, this restriction also make the
fidelity fails to identify those phase transitions that is of
infinite order.

To check this point, also verify the above expression, we now study
the fidelity susceptibility of a non-trivial model in condensed
matter physics, more specifically, the one-dimensional Hubbard
model, where the phase transition happened at the half-filling case
is of infinite order. The Hamiltonian of the Hubbard model reads
\begin{equation}
H=-t\sum_{\langle jl\rangle \sigma} c^\dagger_{j,\sigma}c_{l,
\sigma}+U\sum_{j}n_{j,\uparrow }n_{j,\downarrow }, \label{eq:HamiltonianHub}
\end{equation}
where $c^\dagger_{j,\sigma}$ and $c_{j,\sigma}$ are the creation and
annihilation operators for electrons with spin $\sigma$ (with
$\sigma=\uparrow,\downarrow$) at site $j$ respectively,
$n_{j,\sigma}=c_{j,\sigma}^\dagger c_{j,\sigma}$, $t$ is the hopping
integral, and $U$ denotes the strength of on-site interaction. At
half filling, the ground state of the Hubbard model undergoes a
quantum phase transition from an ideal conductor to a Mott-insulator
at the point $U=0$ \cite{EHLieb68}. For simplicity, we diagonalize
the Hamiltonian for a 10-site systems with periodic boundary
conditions using Lanczos method, and compute fidelities with various
interaction intervals and their corresponding susceptibility.
Numerical results are shown in Fig. \ref{fig:fid_hub}. Obviously,
they support our conclusion that fidelity susceptibility rather than
fidelity is more crucial in the ground state. That is the fidelity
susceptibility does not depend on the value of $\delta U$, but the
fidelity does. This fact makes it possible to evaluate the fidelity
from the ground state $|\Psi_0(U)\rangle$ without knowledge of
$|\Psi_0(U\pm\delta U/2)\rangle$.

Another interesting observation is that neither fidelity
susceptibility is a maximum, nor fidelity is a minimum at the
critical point. This fact goes somewhat beyond the physical
intuition aroused by the original research motivation
\cite{Pzanardi2006} which expects that the fidelity should be a
minimum at the critical point. Though finite-size scaling analysis
(see Fig. \ref{fig:fid_hubscale}) shows that the fidelity
susceptibility at $U=0$ may become larger and larger as the system
size increases, the divergence of the rescaled fidelity
susceptibility $\chi_F/N$ \cite{Notesadded} at the critical point is
an unexpected phenomenon (See Fig. \ref{fig:chi_hub}). It is because
for the half-filled Hubbard model, the phase transition at $U=0$ is
of the Kosterlitz-Thouless type. In this case, the ground-state
energy as a function of $U$ can be expanded to any order around
$U=0$, and the density-density correlation defined by the $U$ term
in the Hamiltonian (\ref{eq:HamiltonianHub}) does not have a
long-range order, then the local order parameter for the
symmetry-breaking theory is not well-defined. On the other hand, if
$U$ deviates from the critical point, the scaling behavior shown in
Fig. \ref{fig:fid_hubscale} manifests that the fidelity
susceptibility may increases in the small $U$ region. However, as
the $L$ increases, the trend of $\chi_F$ is still similar to the
case of $U=0$. Therefore, to find an appropriate driving term is
very crucial for understanding the role of fidelity in quantum
critical phenomena. For the Hubbard model, besides the $U$ term, it
seems be very necessary to introduce another non-local term for the
fidelity. Nevertheless, it is still a challenging problem at
present.

\section{Mixed-state fidelity and thermal phase transitions}

The generalization of fidelity to finite temperatures is proposed recently.
Based on the definition of fidelity between two mixed states, it has been shown
that fidelity can be expressed in terms of the partition function \cite{Pzanardi0612006},

\begin{eqnarray}
F_i(\beta, \delta)=\frac{Z(\beta)}{\sqrt{Z(\beta -\delta\beta/2) Z(\beta
+\delta\beta/2)}},
\end{eqnarray}
where $\beta=1/T$, and
\begin{eqnarray}
Z(\beta)=\sum_{n}e^{-\beta E_n} =\sum_E g(E) e^{-\beta
E}.\label{eq:partitionfun}
\end{eqnarray}
Here $g(E)$ is the density of states and can be calculated from Monte-Carlo
simulations \cite{DPLandaub} such as Wang-Landau algorithm \cite{Fwang2001}.
Then the fidelity susceptibility driven by temperature can be calculated as
\begin{eqnarray}
\chi_F=\left.\frac{-2\ln F_i}{\delta\beta^2}\right|_{\delta\beta\rightarrow 0}
=\frac{C_v}{4\beta^2}.
\end{eqnarray}
Similarly, if the driving term in the Hamiltonian is a Zeemann-like term, which
is crucial in Landau's symmetry-breaking theory,  then the fidelity
susceptibility is simply the magnetic susceptibility $\chi$,
\begin{eqnarray}
\chi_F=\left.\frac{-2\ln F_i}{\delta h^2}\right| _{\delta h\rightarrow 0} =
\frac{\beta \chi}{4}.
\end{eqnarray}
Clearly, the specific heat is simply the fluctuation of the internal energy,
i.e. $C_v=\beta^2(\langle E^2\rangle -\langle E\rangle^2)$, while the magnetic
susceptibility is the fluctuation of the magnetization, i.e.
$\chi=\beta(\langle M^2\rangle -\langle M\rangle^2)$. Thus fidelity
susceptibility is just the fluctuation (structure factor) of the driving term
in the Hamiltonian.

\section{Summary}

In summary, we established a general relation between the fidelity and the
structure factor of the driving term of the Hamiltonian for both quantum and
classical critical phenomena. Such relation not only enables us to evaluate
fidelity easily through prevailing numerical techniques such as DMRG, ED,
and Monte-Carlo simulations, but also builds a straightforward connection
between the concepts in quantum information theory and those in quantum
many-body physics.

We thank X. G. Wen and H. Q. Lin for helpful discussions. This work is supported
by the Earmarked Grant for Research from the Research Grants Council of HKSAR,
China (Project CUHK N\_CUHK204/05 and 400906).


\begin{thebibliography}{99}
\bibitem{Notesadded}
Divergence of $\chi$ is meaningless. Since the value of $\delta U$ should depend
on the system size, otherwise, the fidelity between any two states separated by
a given $\delta U$ usually is zero for an infinite system.
\end{thebibliography}
\end{document}